*Classic hydrodynamic and kinetic formalism as averaging of delta-functional particle images*
L. S. Kuz'menkov, and P. A. Andreev,
Lomonosov Moscow State University, Russia

**Abstract**
Critical analyses of well-known methods of derivation of kinetic and hydrodynamic equations is presented. Another method of derivation of kinetic and hydrodynamic equations from classic mechanics is described. It is shown that equations of classic hydrodynamics can be derived directly from microscopic picture of motion, without using of kinetic equations as an intermediate step. New method of derivation of equation of macroscopic motion includes explicit averaging of microscopic motion on infinitesimally small piece of medium. This averaging leads to presence of electric dipole, magnetic dipole, and higher moments along with the charge density and the current density in hydrodynamic equations. The method under consideration allows to derive equations of evolution for new quantities.

Since in 1938, A. A. Vlasov suggested his famous kinetic equation, problem of derivation of this equation from microscopic picture of classical particle motion were appeared. A method of derivation were suggested by N. N. Bogoliubov (Bogoliubov–Born–Green–Kirkwood–Yvon hierarchy). They used integration of the Liouville equation over coordinates and momentums of all particles except one. It gave equation for "one-particle" distribution equation [1]. Which describes evolution of many-particle systems in six-dimensional phase space. However we should keep in mind that the Liouville equation constructed of dynamical functions, such as coordinate of a particle $\mathbf{r}_i(t)$ and momentum of a particle $\mathbf{p}_i(t)$, which are functions of time corresponding to equations of motion. Thus the Liouville equation is also an equation for a dynamical function. Integrating this equation we obtain equation for reduced dynamical function $f = f(\mathbf{r}(t), \mathbf{p}(t), t)$ instead of six-dimensional scalar field for true distribution function $f = f(\mathbf{r}, \mathbf{p}, t)$, where $\mathbf{r}$, $\mathbf{p}$, $t$ are independent variables. Consequently, the method of Bogoliubov chains does not give us kinetic equations, but structures with similar form, but another mathematical structure. It shows us that problem of derivation of kinetic equations were not solved.

Next important step in construction of classic theory of many-particle motion were made using the Dirac delta function for representation of Newton dynamics from configurational space to three dimensional physical space [1], [2]. Since particles in classic theory are considered as point like objects each particle can be described as the Delta function image $\delta = \delta(\mathbf{r} - \mathbf{r}(t))$, where $\mathbf{r}$ presents a point in space and $\mathbf{r}(t)$ gives position of a particle due to its' evolution governed by the Newton equation. Making sum of delta functions over all particles in system under consideration we obtain microscopic concentration of particles [1],[2]. Deriving time evolution of the microscopic concentration we get continuity equation, where explicit definition of the particle current appears. Its' definition can be used to derive Euler equation. In the same way we can derive equations of evolution of other collective physical variables. These equations describe collective microscopic evolution of classical particles. They have form of hydrodynamic equations. Nevertheless if we want to get macroscopic hydrodynamic equations we should average these equations over infinitesimally small piece of medium. Formal averaging can give us a part of full macroscopic picture. However it would be better to have explicit form of averaging.

We get used to derive hydrodynamic equations from kinetic equations [1]. However, we can see that hydrodynamic equations can be derived directly from microscopic evolution of particles [3].

Similarly to concentration of particles in physical three dimensional space, the concentration of particles in six-dimensional phase space, which is the distribution function, were defined (see for instance [1]). This distribution function allows to derive equations of microscopic motion in terms of another set of many-particle function having structure of kinetic equations. Which has to be averaged to give equations of macroscopic evolution.

Averaging is getting us from microscopic level of separated particles to macroscopic level of a medium.

Here we discuss a method of averaging suggested and developed in Refs. [3], [4], which can be used for independent derivation of different branches of macroscopic theory: Hydrodynamics and Kinetics.

Let us consider avenging in three dimensional physical space leading us to macroscopic hydrodynamic equations. Considering microscopic concentration we have infinity in a point where a particle is located and we have zero in points where are no particle. We define averaged concentration considering a point in physical space and considering a vicinity of the point. Next we count number of particles in the vicinity and divide it on volume of the vicinity. Thus we construct the averaged concentration on macroscopic scale in the chosen point at fixed moment of time. Repeating this procedure for all points of three dimensional physical space we find the field of particle concentration in the fix moment of time [5]

$$n(\mathbf{r},t) = \frac{1}{\Delta} \sum_{i=1}^{N(\mathbf{r},t)} 1_i, \quad (1)$$

where $N(\mathbf{r},t)$ is the number of particle in the vicinity constructed around point $\mathbf{r}$ at moment of time $t$, and $\Delta$ is the volume of vicinity. We should mention that the volume of all vicinities is the same. However, formula (1) is not very useful. We do not have information about number of particles in any vicinity. These numbers are determined by microscopic evolution of particles governed by the Newton equations. We need to represent formula (1) in useful form and we need to find an assistant, which will count number of particles in each vicinity. It happens that the Dirac delta function is our assistant. Let us rewrite formula (1) in following form [3], [5]

$$n(\mathbf{r},t) = \frac{1}{\Delta} \int_\Delta d\boldsymbol{\xi} \sum_{i=1}^{N} \delta(\mathbf{r}+\boldsymbol{\xi}-\mathbf{r}_i(t)), \quad (2)$$

where $N$ is the full number of particles in system under consideration, $\mathbf{r}_i(t)$ is the law of motion of $i$-th particle, vector $\boldsymbol{\xi}$ begins in the center of vicinity having coordinate $\mathbf{r}$, vector $\boldsymbol{\xi}$ scanning the vicinity. If a particle is in a vicinity when $\mathbf{r}+\boldsymbol{\xi}-\mathbf{r}_i(t) = 0$. Consequently integral of delta function of $i$-th particle gives 1. If a particle is outside of the vicinity when $\mathbf{r}+\boldsymbol{\xi}-\mathbf{r}_j(t) \neq 0$. Consequently integral of delta function of $j$-th particle equals to zero. Therefore delta function $\delta(\mathbf{r}+\boldsymbol{\xi}-\mathbf{r}_i(t))$ in formula (2) allows to count particles in each vicinity. This delta function keeps information about particle evolution due to microscopic evolution. Differentiating the particle concentration (2) with respect to time we find continuity equation, where an explicit form of the particle current appears. Having the particle current we can derive next equation, which is the Euler equation. Similarly we can derive equations for other collective hydrodynamic variables, for instance the energy density and the thermal pressure tensor.

Similarly we can independently derive kinetic equations. To include picture of particle distribution in momentum space we should repeat reasoning lead us to formula (1), but in six dimensional phase space

$$f(\mathbf{r},\mathbf{p},t) = \frac{1}{\Delta_\mathbf{r}} \frac{1}{\Delta_\mathbf{p}} \sum_{i=1}^{N(\mathbf{r},\mathbf{p},t)} 1_i, \quad (3)$$

where $f(\mathbf{r},\mathbf{p},t)$ is the "one-particle" distribution function, $\Delta_\mathbf{r}$ is the vicinity of a point in coordinate space, $\Delta_\mathbf{p}$ is the vicinity of the very point in momentum space, and $N(\mathbf{r},\mathbf{p},t)$ is the number of particles in six dimensional vicinity of the point.

Again, formula (3) is not useful as formula (1), so we need to engage the Dirac delta functions. In this case we should take two delta functions for each particle, one for coordinate, and the second one for momentum. Finally we have [4]

$$f(\mathbf{r},\mathbf{p},t) = \frac{1}{\Delta_{\mathbf{r}}}\frac{1}{\Delta_{\mathbf{p}}} \int_{\Delta_{\mathbf{r}}} d\xi \int_{\Delta_{\mathbf{p}}} d\eta \sum_{i=1}^{N} \left( \delta(\mathbf{r}+\xi-\mathbf{r}_i(t))\delta(\mathbf{p}+\eta-\mathbf{p}_i(t)) \right), \quad (4)$$

where we use the same notions as in formula (2) along with some extra notions: vector scanning vicinity of the point in momentum space, so for a particle to get in six dimensional vicinity of the point its' coordinate and momentum should satisfy to two relations $\mathbf{r}+\xi-\mathbf{r}_i(t)=0$ $\mathbf{p}+\eta-\mathbf{p}_i(t)=0$, for some $\xi$ and $\eta$ in the vicinity.

Having macroscopic distribution function (4) we can derive kinetic equation for this function taking derivation of the function with respect to time, which will need truncation to get closed model.

It is worthwhile to mention that macroscopically infinitesimally small piece of medium featured not by full charge of particles in vicinity, but also by electric-, magnetic-dipole moments, electric-, magnetic-quadruple moments etc (see Refs. [3], [5]).

To get a set of equations in the five moment approximation we have to derive five equation: equations of number of partivles, momentum and energy evolution. However we have to neglect by polarisation, magnetisation, and higher moments. Thus this approach allows to make different generalizations. Apart from usual thirteen momentum approximation including details of distribution of particles in momentum space via evolution of pressure tensor and energy flux, we can manage other approximations. For instance we can collect polarisation and magnetisation along with the particle concentration, momentum density and energy density [4]. In this way we can trace evolution of electromagnetic field in more details.

Quantum generalization of the many-particle concentration and distribution function we considered in Refs. [6], [7].

Quantum microscopic many-particle concentration appears to be

$$n_Q = \langle \hat{n} \rangle = \int \Psi^* \hat{n} \Psi dR = \int \Psi^*(R,t) \sum_{i=1}^{N} \delta(\mathbf{r}-\mathbf{r}_i) \Psi(R,t) dR. \quad (5)$$

Definition (5) was suggested in Ref. [6] as the first step towards derivation of quantum hydrodynamic equations. Definition (5) appears in condense with general rules of quantum mechanics. We took classic microscopic concentration of particles $n = \sum_{i=1}^{N} \delta(\mathbf{r}-\mathbf{r}_i(t))$, quantized it replacing coordinates of particles by the operator of coordinate. This procedure gave us operator of particle concentration. To get function of concentration we have to take quantum mechanical average of the operator and find formula (5) [6].

Similar steps give us quantum distribution function

$$f_Q = \langle \hat{f} \rangle = \int \Psi^*(R,t) \left( \sum_i \delta(\mathbf{r}-\mathbf{r}_i)\delta(\mathbf{p}-\hat{\mathbf{p}}_i) \right) \Psi(R,t) dR.$$

For more details on quantum distribution function see Ref. [7].

We have shown that equations of classic hydrodynamics can be directly derived from microscopic evolution with no intermediate steps as kinetic equations. We have found that getting on macroscopic level leads to arising of contribution of polarization, magnetization, density of electric quadruple moment and etc. We demonstrate relation of this method with the many-particle quantum hydrodynamics.

### References

1. Yu. L. Klimontovich, *Statistical Physics [in Russian]*, Nauka, Moscow (1982); English transl., Harwood, New York (1986).

2. S. Weinberg, *Gravitation and Cosmology* (John Wiley and Sons, Inc., New York, 1972).
3. M. A. Drofa and L. S. Kuz'menkov, *Continual approach to multiparticle systems with long-range interaction. Hierarchy of macroscopic fields and physical consequences*, Theor. Math. Phys. **108,** 849 (1996).
4. L. S. Kuz'menkov, *Field form of dynamics and statistics of systems of particles with electromagnetic interaction*, Theor. Math. Phys., **86,** 159 (1991).
5. L. S. Kuz'menkov and P. A. Andreev, *Microscopic Classic Hydrodynamic and Methods of Averaging*, PIERS Proceedings, Moscow, Russia, August 19-23, p. 158, 2012.
6. L. S. Kuz'menkov, S. G. Maksimov, *Quantum hydrodynamics of particle systems with coulomb interaction and quantum Bohm potential*, Theor. Math. Phys., **118**, 287 (1999).
7. P. A. Andreev, *Quantum kinetics of spinning neutral particles: General theory and Spin wave dispersion*, arXiv: 1308.3715.